\documentclass[traditabstract,letter]{aa} % for the abstract without structuration 
\usepackage{graphicx}
\usepackage{txfonts}
\begin{document}
   \title{\emph {Herschel}\thanks{\emph {Herschel} is an ESA space observatory with science instruments provided by Principal Investigator consortia. It is open for proposals for observing time from the worldwide astronomical community.} PACS Spectroscopic Diagnostics of Local ULIRGs:  Conditions and Kinematics in Mrk 231}

 %  \subtitle{I. Overviewing the $\kappa$-mechanism}

   \author{J. Fischer
          \inst{1}\fnmsep\thanks{Visiting scientist at Max-Planck-Institut f\"{u}r extraterrestrische Physik (MPE), Garching, Germany}
          \and
          E. Sturm\inst{2}
          \and
          E. Gonz\'{a}lez-Alfonso\inst{3}
          \and
          J. Graci\'{a}-Carpio\inst{2}
          \and
          S. Hailey-Dunsheath\inst{2}
          \and
          A. Poglitsch\inst{2}
          \and
          A. Contursi\inst{2}
          \and
          D. Lutz\inst{2}
          \and
          R. Genzel\inst{2}
          \and
          A. Sternberg\inst{4}
          \and
          A. Verma\inst{5}
          \and
          L. Tacconi\inst{2}
          }

  \institute {Naval Research Laboratory, Remote Sensing Division, 4555 Overlook Ave SW, Washington, DC 20375, USA\\
  	\email{jackie.fischer@nrl.navy.mil}
   	\and
             	Max-Planck-Institut f\"{u}r extraterrestrische Physik (MPE), Postfach 1312, D-85741 Garching, Germany 
         \and
         		Universidad de Alcala de Henares, Departamento de Fisica, Campus Universitario, Spain
	\and
		Sackler School of Physics and Astronomy, Tel Aviv University, Tel Aviv 69978, Israel
	\and
		University of Oxford, Denys Wilkinson Building, Keble Road, Oxford OX1 3RH, United Kingdom	
               }
  %       \and
  %       University of Alexandria, Department of Geography, ...\\
  %           \email{c.ptolemy@hipparch.uheaven.space}
  %          \thanks{The university of heaven temporarily does not
  %                  accept e-mails}

   \date{Received 31 March 2010; accepted 5 May 2010}

% \abstract{}{}{}{}{} 
% 5 {} token are mandatory
 
  \abstract
   {In this first paper on the results of our \emph {Herschel} PACS survey of local Ultraluminous Infrared Galaxies (ULIRGs), as part of our \emph {SHINING} survey of local galaxies, we present far-infrared spectroscopy of Mrk 231, the most luminous of the local ULIRGs, and a type 1 broad absorption line AGN.  For the first time in a ULIRG, all observed far-infrared fine-structure lines in the PACS range were detected and \emph{all} were found to be deficient relative to the far infrared luminosity by 1 -- 2 orders of magnitude compared with lower luminosity galaxies.  The deficits are similar to those for the mid-infrared lines, with the most deficient lines showing high ionization potentials.  Aged starbursts may account for part of the deficits, but partial covering of the highest excitation AGN powered regions may explain the remaining line deficits.  A massive molecular outflow, discovered in OH and $^{18}$OH, showing outflow velocities out to at least 1400 km sec$^{-1}$, is a unique signature of the clearing out of the molecular disk that formed by dissipative collapse during the merger.  The outflow is characterized by extremely high ratios of $^{18}$O / $^{16}$O  suggestive of interstellar medium processing by advanced starbursts.     
    }
   \keywords{infrared: galaxies -- galaxies: ISM -- quasars: absorption lines -- galaxies: individual: Mrk 231
               }
  \authorrunning {J. Fischer et al. } 
  \titlerunning {\emph {Herschel} PACS Far-infrared Spectroscopy of Mrk 231}
   \maketitle
%
%________________________________________________________________

\section{Introduction}
ULIRGs (L$_{IR}$ $\geq 10^{12} L_{\odot}$), which in the local universe are known to be mergers of gas rich galaxies, are thought to play a critical role in galaxy evolution.  It is therefore crucial to understand the conditions, dynamics, chemistry and energetics of this stage of evolution, in which dissipative collapse accompanies the transformation of gas rich galaxies into ellipticals (Lonsdale et al. 2006). As part of the \emph{SHINING} Key Project on local galaxies, we are carrying out a far-infrared (FIR) spectroscopic survey of all 21 ULIRGs in the Revised Bright Galaxy Survey (Sanders et al. 2003).  Previous work on ULIRGs with the Infrared Space Observatory (ISO) Long Wavelength Spectrometer (LWS) found deficits in far-infrared atomic and ionized fine-structure line emission relative to their FIR luminosities (Luhman et al. \cite{Luhman98}) accompanied by prominent molecular absorption in excited transitions of molecules such as OH and H$_2$O rarely seen in other galaxies (Fischer et al. \cite{Fischer99}).   The deficits, often based only on a few non-detections, were postulated to result from the high ratios of UV radiation density to particle density in the nuclei of these galaxies (Malhotra et al. 2001, Luhman et al. 2003, Abel et al. 2009), high gas density (e.g. Negishi et al. 2001)  and high FIR opacity and/or high luminosity-to-mass ratio (Gonz\'{a}lez-Alfonso et al. 2004, 2008, hereafter GA04, GA08), but because of the paucity of diagnostic line detections this puzzle remained unresolved.  GA04 and GA08 showed that the high excitation molecular absorptions in both Arp 220 and Mrk 231 are radiatively pumped, consistent with high radiation density to particle density. They identified far-infrared absorption by OH, H$_2$O, NH$_3$, NH, and $^{18}$OH, but the identifications of the latter species were uncertain due to the low resolution of the LWS.  The PACS ULIRG spectroscopic survey is designed to kinematically identify the ionized, atomic and molecular regions and to study the conditions of the interstellar medium, thereby illuminating the nature of this important evolutionary phase.  In this first paper we present our initial results on Mrk 231, the most luminous of the local ULIRGs (L(8--1000 $\mu$m) = 3.2 $\times$ 10$^{12}$ L$_{\odot}$) and a type 1, low-ionization broad absorption line (LoBAL) active galactic nucleus (AGN) at an adopted distance of 172 Mpc (z=0.04217).  Its central quasar is covered by a semi-transparent dusty shroud producing about 3.1 magnitudes of extinction at 4400 $\AA$ (Reynolds et al. 2009) and is at the center of a rotating, nearly face-on molecular disk (Downes \& Solomon 1998).  Based primarily on \emph {Spitzer} results, Veilleux et al. (2009) estimate that the average AGN contribution to the bolometric luminosity in ULIRGs is 35 -- 40\% and that for Mrk 231 the AGN contribution is $\sim$ 70\% by most estimation techniques.  The contribution of an advanced 120 -- 250 Myr nuclear starburst is estimated at 25 -- 40\% based on near infrared observations of Mrk 231 by Davies et al. (2007).  

%__________________________________________________________________
\begin{table*}
\caption{Spectroscopic measurements for Mrk 231}             
\label{Table1}      
\centering          
\begin{tabular}{r ccccc}     % 6 columns 
\hline\hline       
                      % To combine 4 columns into a single one 
Transition\tablefootmark{a,}\tablefootmark{b} & Line flux  & Line flux error\tablefootmark{c} & Equivalent Width & Measured FWHM & Inferred FWHM\tablefootmark{d}\\ 
 & 10$^{-17}$ W m$^{-2}$	 & 10$^{-17}$ W m$^{-2}$	& 10$^{-3}$ $\mu$m	& km s$^{-1}$	& km s$^{-1}$ \\
\hline
                    
[N III] 57.3 $\mu$m & 2.8  & 0.6 & -1.0 & 203 & 177\\
  
[O I] 63.2 $\mu$m & 36.0 & 2.6 & -17.0 & 232 & 218\\  

[O III] 88.4 $\mu$m & 4.1 & 0.6 & -3.8 & 311 & 289\\

[N II] 121.9 $\mu$m & 4.1 & 0.5 & -10.4 & 389 & 266\\

[O I] 145.5 $\mu$m & 3.2 & 0.4 & -17.9 & 324 & 208\\

[C II] 157.7 $\mu$m & 38.3 & 1.3 & -296.0 & 337 & 247\\

[N II] 205.2 $\mu$m\tablefootmark{e} & 2.6 & 0.2 & -61.0 & 524 & 242\\

OH 79.1, 79.2 $\mu$m & -22.8 & 1.0 & 15.3 & -- & --\\
$ $ & 3.2 & 0.6 & -2.1 & -- & --\\

OH 119.2, 119.4 $\mu$m & - 47.6 & 0.3 & 107.4 & -- & --\\
$ $ & 9.4 & 0.3 & -21.6 & -- & --\\

$^{18}$OH 120.0, 120.2 $\mu$m & -14.7 & 0.2 & 34.3 & -- & --\\
$ $ & 1.7 & 0.2 & -4.0 & -- & --\\

H$_2$O 78.7 $\mu$m & -18.0 & 0.1 & 12.2 & 444 & 422\\

HF $\diagup$ H$_2$O 121.7 $\mu$m & -4.7 & 0.4 & 11.8 & 560 & 483\\

\hline                  
\end{tabular}\\
\tablefoottext{a}{For doublets, fluxes listed are for the blended pair.}
\tablefoottext{b}{For transitions with both absorption and emission, the absorption component is listed first.}
\tablefoottext{c}{Errors listed are the statistical rms uncertainties;  we allow 25\% calibration uncertainties (see text).}
\tablefoottext{d}{Inferred velocity width is based on the instrument resolution assuming a Gaussian profile;  FWHMs are not listed for doublets.}
\tablefoottext{e}{{Herschel} SPIRE data (V10).}
\end{table*}

%__________________________________________________________________

\section{Observations and data reduction}

The observations were taken with the Photodetector Array Camera and Spectrometer (PACS) integral field spectrometer (Poglitsch et al. 2010) on board the \emph {Herschel} Space Observatory (Pilbratt et al. 2010) in high spectral sampling range spectroscopy mode using small chop-nod cycles. Scans covering the range $\pm$ 1300 km s$^{-1}$ were taken around five fine-structure lines and a longer range scan included the OH 119 $\mu$m $^{2}\Pi_{3/2}$ 5/2 -- 3/2 $\Lambda-$doublet transitions, the $^{18}$OH 120 $\mu$m counterparts, and the [NII] 122 $\mu$m fine-structure line.  The OH 79 $\mu$m $^{2}\Pi_{1/2}$ -- $^{2}\Pi_{3/2}$ 1/2 -- 3/2 $\Lambda-$doublet transitions, the nearly superposed $^{18}$OH counterparts, and the H$_2$O 78.7 $\mu$m 4$_{23}$ -- 3$_{12}$ line were observed in parallel on the blue array.  

The basic data reduction was done using the standard PACS reduction and calibration pipeline (ipipe) included in HIPE 2.0 1340\footnote{HIPE is a joint development by the \emph {Herschel} Science Ground Segment Consortium, consisting of ESA, the NASA \emph {Herschel} Science Center, and the HIFI, PACS and SPIRE consortia.}.  The data are consistent with an unresolved point source.  For a well centered point source, the peak line and continuum flux will fall on the central of the PACS 5 x 5 spaxels (9$\arcsec$ x 9$\arcsec$ spatial pixels), with nearly all of the rest of the flux falling on the remaining pixels of the array.  The best signal--to--noise on a faint line is obtained by scaling the line profile on the central spaxel, to the continuum flux integrated over the array.  Where better baselines were obtained by scaling the sum of the brightest spaxels, this was done instead.  For the [CII]158 $\mu$m scan, all of the spaxels were co-added.  

\begin{figure}
   \centering
   \includegraphics[width=5.7cm]{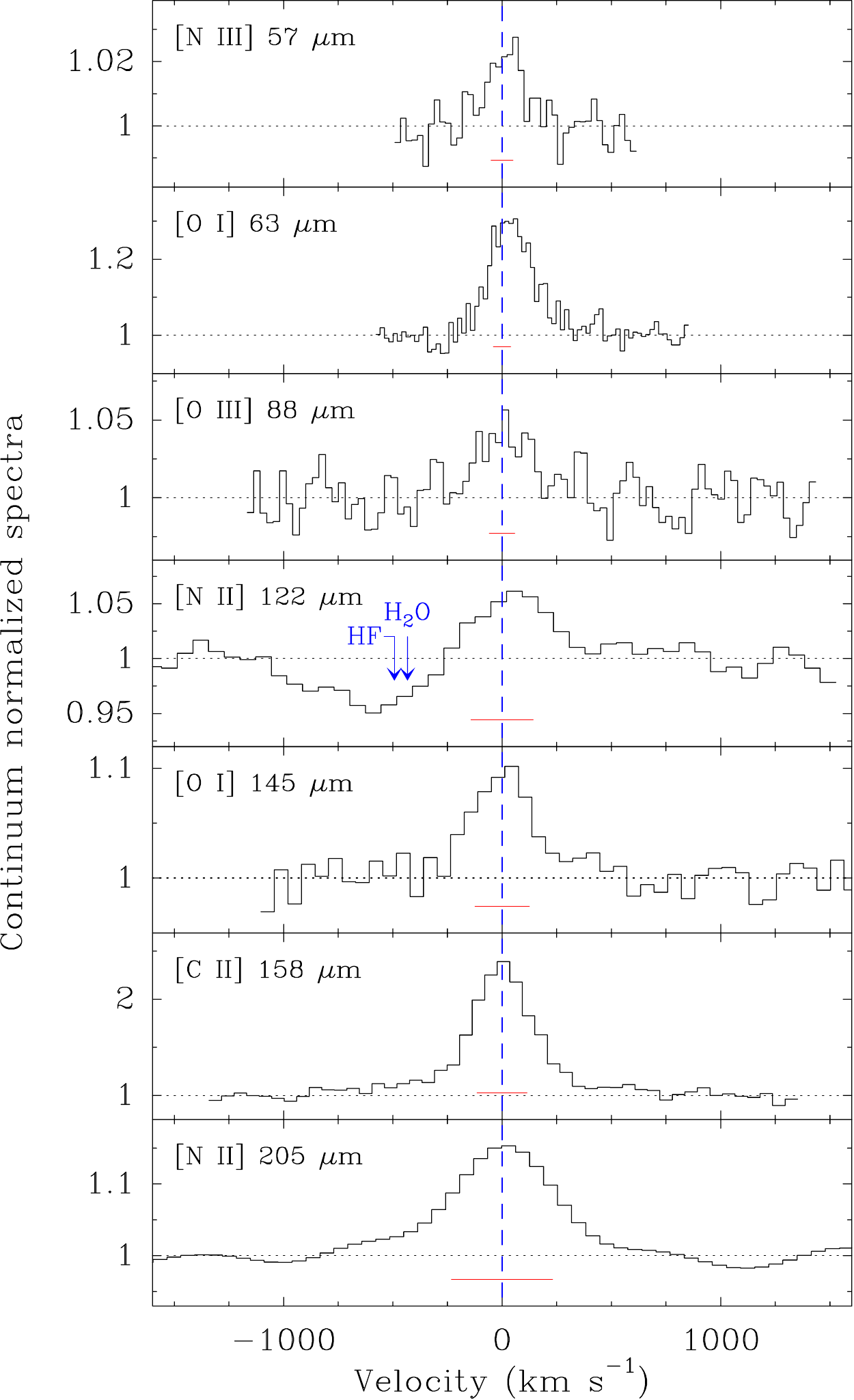}
      \caption{Mrk 231 normalized far-infrared fine-structure line profiles versus velocity.  The red bars represent the instrumental FWHMs.  The absorption observed to the blue side of the [NII]122 $\mu$m line is due to the HF 2-1 and/or H$_2$O 4$_{32}$ -- $4_{23}$ lines. 
              }
         \label{FineStructureLines}
   \end{figure}

At the time of preparation of this paper the full planned calibration procedure is not yet incorporated into the pipeline.  For calibration of the PACS line fluxes, the line scan continua were therefore scaled to a fit to existing photometry of Mrk 231.  The integrated line fluxes, statistical errors, equivalent widths, measured Gaussian-fit $\Delta$v$_{FWHM}$ and the inferred intrinsic $\Delta$v$_{FWHM}$ obtained by subtracting the instrumental $\Delta$v$_{FWHM}$ in quadrature are listed in Table~\ref{Table1}.   We note that the [CII], [OI]63, and OH119 line fluxes are in agreement with ISO LWS line fluxes reported in Luhman et al. (2003) and GA08 to within the statistical uncertainties, but the stronger [NII] line flux reported in GA08 is not consistent with our observations.  Additionally the [N II] 205 $\mu$m line measured by \emph {Herschel} SPIRE  is included in Table~\ref{Table1} and Fig.~\ref{FineStructureLines} based on measurements presented in van der Werf et al. (2010, hereafter V10).   We estimate the calibration uncertainties to be $\pm$ 25\%.  

  \begin{figure}
   \centering
   \includegraphics[width=8.0cm]{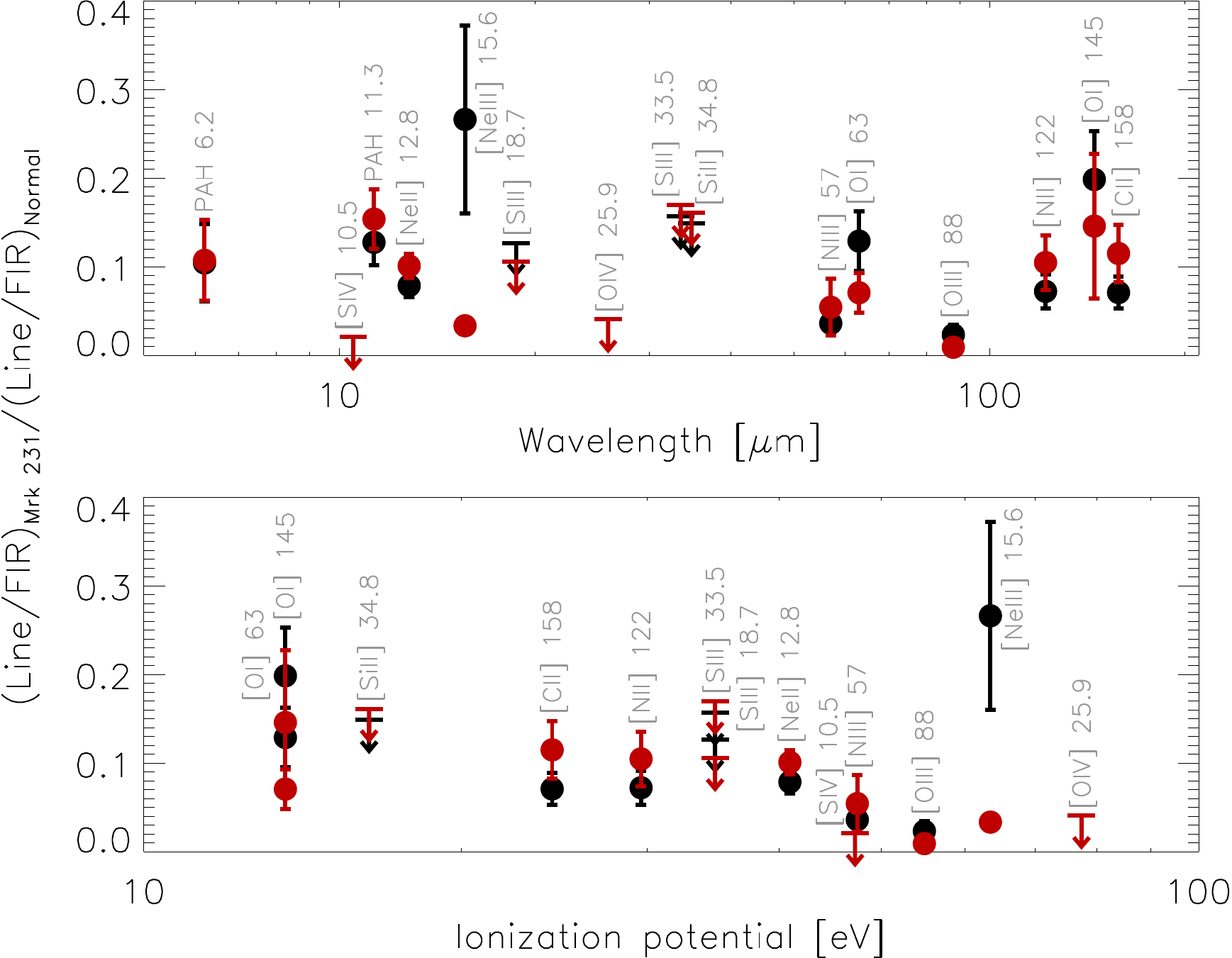}
      \caption{Line-to-FIR ratio in Mrk 231 divided by the median line-to-FIR ratio in a sample of HII galaxies (black symbols) and AGN (red symbols) versus wavelength (\emph{top}) and ionization potential (\emph{bottom}).  The mid-infrared line fluxes in Mrk 231 are from Armus et al. (2007).  The error bars include the Mrk 231 calibration uncertainties and the statistical errors for the comparison sample values.
               }
         \label{LineDeficits}
   \end{figure}

%__________________________________________________________________

\section{Results and Discussion}

\subsection{The fine-structure lines}

The continuum normalized fine-structure line spectra are displayed in Fig.~\ref{FineStructureLines} as a function of velocity, where Mrk 231's redshift of 0.04217 is defined as 0 km s$^{-1}$ and the red horizontal bars represent the instrumental FWHMs.  The FWHM velocities observed in these lines are in all cases larger than the instrumental resolution and we do not detect self-absorption.  The average of the derived intrinsic FWHMs for the fine-structure lines is 235 km s$^{-1}$, similar to the FWHM velocity widths of 167 km s$^{-1}$ and  270 km s$^{-1}$ measured by Sanders et al. (1991) and Tacconi et al. (2002) from CO(1--0) and near-infrared stellar CO bandhead studies, respectively.  Thus the fine-structure line emission is probably associated with the central star forming disk.  In  the [CII] line, wings not characteristic of the instrumental profile are present, out to $-1000$ km s$^{-1}$ for the blue wing, suggesting that some C$^+$ is associated with the outflow discussed in the next subsection.  The [NII]205 profile also suggests a wing out to $-700$ km s$^{-1}$, consistent with the detection of optical [NII] lines by Rupke et al. (2005).  We  estimate that the absorption by HF or H$_{2}$O observed in the blue wing of the [NII] 122 $\mu$m line does not affect the [NII] line flux by more than 50\%, since the FWHM of this transition is similar to those of the other fine-structure lines.  The observed [NII]205 to [NII]122 ratio of 0.6 $\pm$ 0.4 translates into a density range for N$^{+}$ of  log n$_{e}$ = $1.4^ {+0.8}_{-0.4}$ (Rubin et al. 1993).  Photodissociation region (PDR) modeling of [CII], [OI], and FIR fluxes yields UV radiation field and density $G_{0}$ $\approx$ 6 $\times$ 10$^{3}$ and $n$ $\approx$ 500 cm$^{-3}$, at the extreme range of average radiation density per particle density seen in extragalactic nuclei (Sturm et al. 2010, Fig. 4).

The fine-structure line strengths relative to the far-infrared luminosities are weak compared with normal HII galaxies and AGN.  In Fig.~\ref{LineDeficits}, we plot as a function of wavelength (top) and ionization potential (bottom) the fine-structure line to far-infrared flux ratio (FIR, 42 -- 122 $\mu$m) in Mrk 231 relative to the median value for subsamples of HII galaxies (black symbols) and AGN (red symbols) with luminosities 1 x 10$^9$ L$_\odot$ $\leq$ L$_{IR}$ $\leq$ 5 x 10$^{11}$ L$_\odot$ and major isophotal diameters D$_{25}$ $\leq$ 250 arcsecs compiled from the literature and our \emph {SHINING} sample (Graci\'{a}-Carpio et al., in prep.).  There are no [NII] 205 $\mu$m line measurements for the comparison samples, so this line is not included in Fig.~\ref{LineDeficits}.  The observed deficits are severe:  typically an order of magnitude for the neutral and low ionization species and up to two orders of magnitude for the higher ionization lines.  There is no correlation between the deficit and wavelength, and only weak correlation with critical density, so \emph{differential} extinction (with wavelength) does not appear to play a significant role, and density does not play a major role.  We find a strong inverse correlation (correlation coefficient, $-0.78$) with ionization potential when compared with the AGN sample.  A similar inverse correlation is found with the starburst comparison sample only if the [NeIII]15 deficit is treated as an outlier.  Low excitation is consistent with the estimation by Davies et al. (2007) that a 120 -- 250 Myr starburst supplies 25 -- 40 \% of the bolometric luminosity of Mrk 231.  However such an aging starburst can create  a deficit of 40\% at most.  The high relative strength of the [OI] 145 $\mu$m line, with its high excitation lower level, is suggestive of increased radiation density per particle density, as parameterized by the ionization parameter $U$ (Luhman et al. 2003; Abel et al. 2009).  The high ratio [OI] 145 / [CII] 158 $\approx$ 1/11 lies between the predicted line ratio for starburst and AGN at the far-infrared F(60)/F(100) color of Mrk 231 with $U$ $\approx$ 10$^{-2}$ (high $U$) in the dust-bounded models of Abel et al. (2009).  The strong inverse correlation relative to the AGN comparison sample in Fig.~\ref{LineDeficits}  is consistent with partial covering, higher for high ionization lines, of the line emitting region, due for example to a face-on molecular torus, with optically thick clumps.

\begin{figure}
   \centering
   \includegraphics[width=6.2cm]{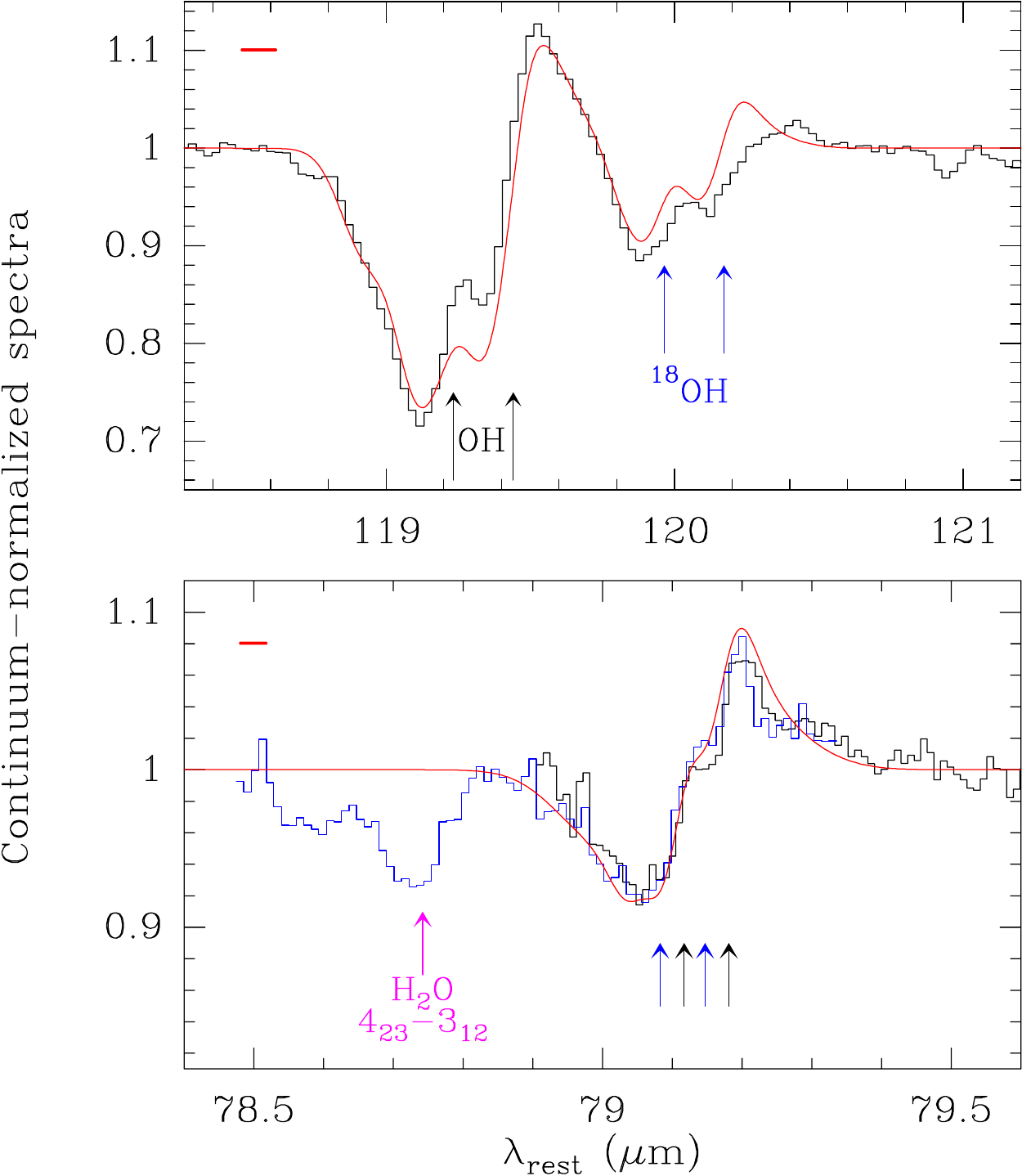}
      \caption{P-Cygni profiles (solid blue, black histograms) of the OH 119 $\mu$m and $^{18}$OH 120 $\mu$m partially resolved doublets (\emph{top}) and the OH 79 $\mu$m unresolved doublet (\emph{bottom}) in Mrk 231 are compared with our modeled profiles (solid red curves, see text).  Absorption in the H$_2$O 4$_{23}$ -- 3$_{12}$ transition with a possible blue-shifted wing is modeled in GA10.  The zero velocity rest-frame wavelengths are marked with black (OH), blue ($^{18}$OH), and magenta (H$_2$O) arrows and the instrumental FWHM is indicated in each panel with horizontal red bars.
              }
         \label{MolecularLines}
   \end{figure}

\subsection{The molecular outflow}

Figure~\ref{MolecularLines} displays the observed 119 -- 120 $\mu$m OH / $^{18}$OH and 79 $\mu$m H$_2$O / OH line profiles in Mrk 231. The OH lines show spectacular P-Cygni profiles in all three ground-state doublets, with extremely broad blue-shifted absorption as far out as 1400 km s$^{-1}$ in the OH 119 $\mu$m line.  Both the high \emph {molecular} outflow velocities and the relative strengths of the OH and $^{18}$OH are unprecedented.  The excited H$_{2}$O 78.7 $\mu$m and HF/H$_{2}$O 121.7 $\mu$m lines (see Fig.~\ref{FineStructureLines}) appear to have blue-shifted wings, indicating that H$_2$O and HF may also be involved in the outflow.  Recently a molecular outflow was detected by Sakamoto et al. (2009) in submillimeter lines of HCO$^+$ and CO toward the nuclei of Arp220 with an outflow velocity of $\sim100$ km s$^{-1}$, but this is modest compared with that of Mrk 231.

Our model of the observed profiles is shown in Fig.~\ref{MolecularLines} and is based on the continuum components and radiative pumping described in Gonz\'{a}lez-Alfonso et al. (2010, hereafter GA10).  It uses three continuum components, hot (radius = 23 pc, $T_{dust}$ = 400 - 150 K), warm (120 pc, 95 K), and extended (610 pc, 41 K), and three velocity components at $100$, $400$, and $600$ km s$^{-1}$ (the latter component with high turbulent velocity, $300$ km s$^{-1}$).  We have conservatively assumed
a screen approach for OH, with the outflow located just around
the warm continuum source, resulting in $N(\mathrm{OH})=1.4\times10^{17}$
cm$^{-2}$. The model indicates mechanical energy
$\gtrsim10^{56}$ erg (dominated by the high velocity component), mechanical
luminosity $\gtrsim1$\% of the TIR luminosity, and an outflow mass of
$\gtrsim7\times10^7$ M$_{\odot}$.   Observations of high-lying OH
lines would help better constrain the outflow location, energetics, and the apparent extremely low $^{16}$O / $^{18}$O ratio.

\begin{figure}
   \centering
   \includegraphics[width=6.8cm]{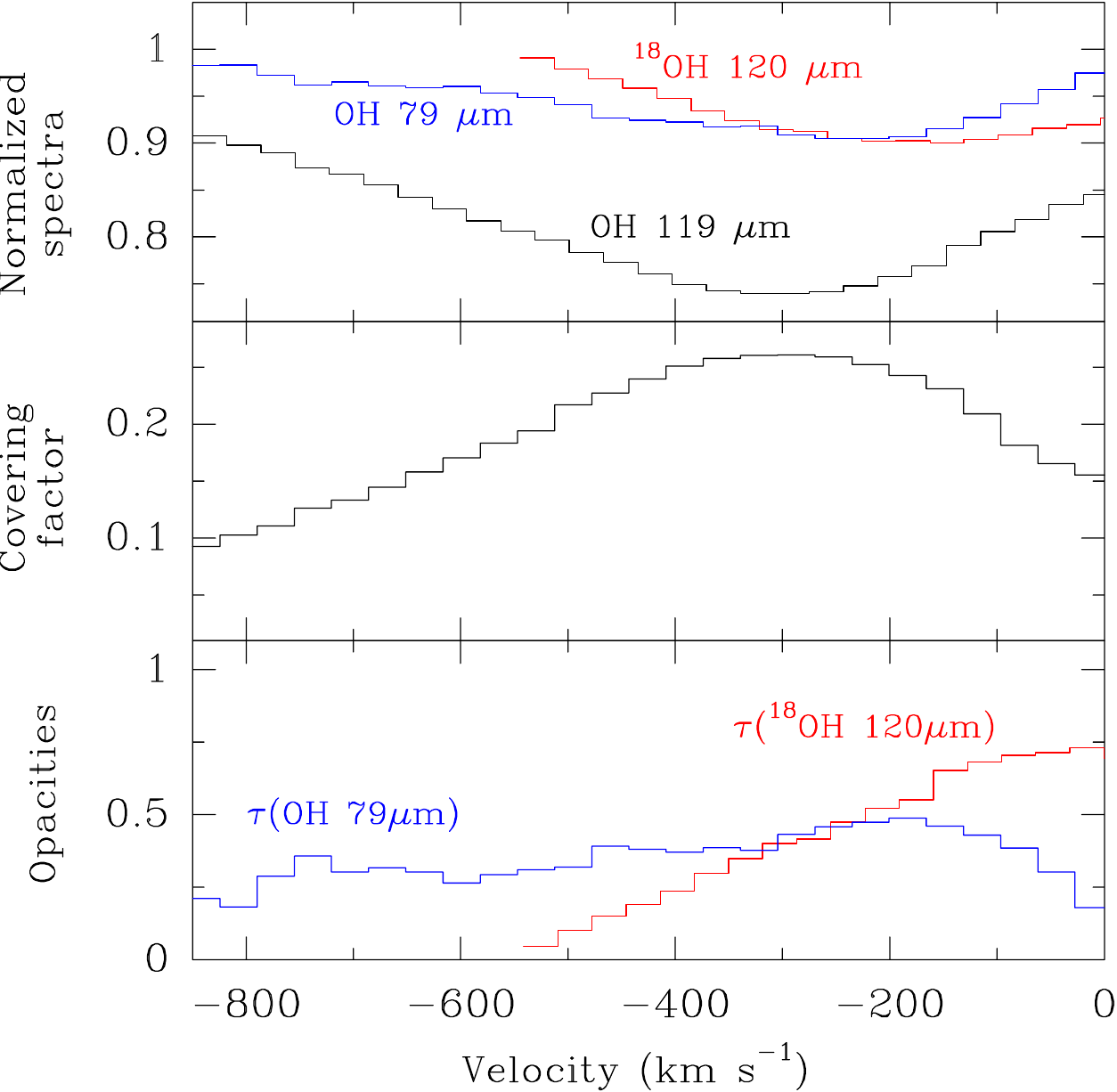}
      \caption{Normalized velocity profiles for the blue-shifted OH and $^{18}$OH absorption (\emph {top}).  The velocity scale is given relative to the bluest component of each OH doublet.  The covering factor (\emph {middle}), assumed to be the same for OH and $^{18}$OH and derived from the profile of the optically thick OH 119 $\mu$m line, is used to calculate the opacities in the 79 $\mu$m and 120 $\mu$m lines (\emph {bottom}, see text). 
              }
         \label{OpticalDepths}
   \end{figure}

Figure~\ref{OpticalDepths} (top) shows that the
absorption at 120 $\mu$m by $^{18}$OH at low outflow velocities ($>-250$ km
s$^{-1}$) is deeper than the corresponding absorption in the
cross-ladder 79 $\mu$m OH doublet at the same velocity.   In this simple model we assume that the incomplete absorption in the optically thick OH 119
$\mu$m doublet is  a result of the fact that OH at a given velocity is only covering a fraction
of the continuum (Fig.~\ref{OpticalDepths}, middle) and we ignore any differential 
re-emission of the $^{18}$OH and 79 $\mu$m OH lines.   We infer that at velocities $>-250$ km s$^{-1}$
$\tau_v(\mathrm{^{18}OH \, 120 \mu m}) > \tau_v(\mathrm{OH \, 79
\mu m})$ (Fig.~\ref{OpticalDepths}, bottom).  From the ratio of the Einstein B coefficients, we find that the opacity in the 79 $\mu$m OH line is 40
times lower than in the 119 $\mu$m OH line and thus that
$N(\mathrm{OH})/N(\mathrm{^{18}OH}) < 40$.  These results are
not critically dependent on the uncertainties in the covering factor due to
re-emission (or thermal emission) in the 119 $\mu$m OH line, which would affect
both the opacities of the $^{18}$OH and 79 $\mu$m OH lines but not their
relationship.  We note that the effects of filling in by cooler dust would weaken the 120 $\mu$m absorption more than the 79 $\mu$m absorption.  The rarer isotopologue $^{17}$OH would appear between the two lines at 119 and 120 $\mu$m, but it is not possible to put a limit on its strength due to the broad width of the detected features.  

We note that CH$^+$, HCl, H$^{37}$Cl, H$^{79}$Br,
and H$^{81}$Br have transitions at $\approx120$ $\mu$m, but HCl and HBr
and their isotopologues are not expected to significantly contaminate the observed
absorption due to the high lower level energies of their $120$ $\mu$m
transitions ($180-240$ K). On the other hand, CH$^+$(3-2) at 119.86 $\mu$m 
could significantly contaminate the blue-shifted component of the
$^{18}$OH doublet, given that CH$^+$(1-0) is detected in Mrk 231 (V10).  We estimate that
contamination by CH$^+$ could decrease the implied $^{18}$OH abundance
by a factor of $1.5-2$ and the $^{16}$O / $^{18}$O ratio could be increased to $\sim$ $30-40$, still significantly below the solar system isotopic ratio of 500, the galactic center ratio of 250, the starburst galaxy ratio of $\sim$ 150, and very close to the enhancement to $\sim$ 50 that is theoretically predicted with a top heavy initial mass function in an advanced starburst stage (Henkel \& Mauersberger 1993).  If our simple model is correct, the high $^{18}$O / $^{16}$O ratio seen here provides stringent constraints for stellar nucleosynthesis models.

Rupke et al. (2005) find that although both AGN and starburst galaxies have massive outflows traced  by broad interstellar Na I D absorption lines, the high velocity ($\Delta$v$\geq$ 2000 km s$^{-1}$) outflows seen almost exclusively in type 1 ULIRGs including Mrk 231, are probably powered by the AGN.  This was also shown in models by Narayanan et al. (2008).  If it is also true for the massive and energetic moderate velocity outflow seen here for the first time in OH and $^{18}$OH, then an AGN driven outflow is in the process of dispersing into the intergalactic medium, highly processed  material produced by an advanced starburst.  The molecular outflow shows velocities similar to the kiloparsec scale outflow component of Mrk 231 seen by Rupke et al. and appears to be a unique signature of the clearing out of the molecular disk that formed with the dissipative collapse during the merger.

%______________________________________________________________

%_____________________________________________________________

\begin{acknowledgements}
PACS has been developed by a consortium of institutes led by MPE 
(Germany) and including UVIE (Austria); KU Leuven, CSL, IMEC (Belgium); 
CEA, LAM (France); MPIA (Germany); INAF-IFSI/OAA/OAP/OAT, LENS, SISSA
(Italy); IAC (Spain). This development has been supported by the funding 
agencies BMVIT (Austria), ESA-PRODEX (Belgium), CEA/CNES (France), DLR 
(Germany), ASI/INAF (Italy), and CICYT/MCYT (Spain).  We thank the HerCULES Key Project for providing the [NII] 205 $\mu$m profile.  Basic research in IR astronomy at NRL is funded by the US ONR.  JF acknowledges support from the NHSC and wishes to thank MPE for its hospitality during the first exciting year of \emph {Herschel} science.  

\end{acknowledgements}

\end{document}